# Current-Controlled Skyrmion Number in Confined Ferromagnetic Nanostripes

*Jialiang Jiang, Jin Tang\*, Yaodong Wu, Qi Zhang, Yihao Wang, Junbo Li, Yimin Xiong, Lingyao Kong, Shouguo Wang, Mingliang Tian, and Haifeng Du\**

Jialiang Jiang, Yaodong Wu, Yihao Wang, Junbo Li, Mingliang Tian, Haifeng Du

Anhui Province Key Laboratory of Condensed Matter Physics at Extreme Conditions, High Magnetic Field Laboratory, Hefei Institutes of Physical Science, Anhui, Chinese Academy of Sciences, Hefei 230031, China

Email: duhf@hmfl.ac.cn

Jialiang Jiang, Yihao Wang, Junbo Li

University of Science and Technology of China, Hefei 230026, China

Jin Tang, Qi Zhang, Yimin Xiong, Lingyao Kong, Mingliang Tian

School of Physics and Optoelectronics Engineering Science, Anhui University, Hefei, 230601, China

Email: jintang@ahu.edu.cn

Shouguo Wang

School of Materials Science and Engineering, Anhui University, Hefei 230601, China

# Abstract

Skyrmions are vortex-like localized magnetic structures that possess an integer-valued topological index known as the skyrmion number or topological charge. Skyrmion number determines the topology-related emergent magnetism, which is highly desirable for advanced storage and computing devices. In order to achieve device functions, it is necessary to manipulate the skyrmion number in confined nanostructured geometries using electrical methods. Here, we report the reliable current-controlled operations for manipulating the skyrmion number through reversible topological transformations between skyrmion chains and stripe domains in confined $Fe_3Sn_2$ nanostripes. The results of micromagnetic simulations are successful in numerically reproducing our experiments and explaining them through the combined effect of current-induced Joule heating and magnetic hysteresis. These findings hold the potential to advance the development of topological spintronic devices.

## 1. Introduction

Magnetic skyrmion, a spin texture with the nanometer scale, is a promising candidate for the next generation of spintronic devices due to its topological properties, including small size, stability, and unique electromagnetism[1-11]. The topology of skyrmions is determined by the skyrmion number $N_s$ (also called winding number or topological charge),[10] defined as $N_s = -1/(4\pi) \int \mathbf{m} \cdot (\frac{\partial \mathbf{m}}{\partial x} \times \frac{\partial \mathbf{m}}{\partial y}) \mathrm{d}x\mathrm{d}y$. Skyrmion number $N_s$ counts how many times the magnetization vector field $\mathbf{m}$ wraps around a unit sphere.[10] It plays a significant role in determining the topology-related properties of skyrmions, such as skyrmion Hall effects[12-14] and read-out electrical transport properties.[15] Recent studies have shown that the topology-dependence of dynamic motions;[16, 17] the Hall effects,[18-21] Nernst effects,[22-24] and anisotropic magnetoresistance[25] are all proportional to the skyrmion number.

It is essential to manipulate the skyrmion number in confined nanostructures to develop topological spintronic devices, e.g. random access memory[26, 27] and neuromorphic computing.[21] Many studies have demonstrated static stabilizations of skyrmions in confined nanostructures, such as disks[28-30] and stripes,[31-33] but electrical manipulations of skyrmion numbers in these geometries are still less explored. Electrical methods mainly include voltage applied to insulators and current applied to metals. Recent research indicates that voltage-controlled strain may be a viable approach to controlling skyrmion numbers in confined disks.[34] Recent studies have demonstrated the current-induced creation of skyrmions from stripe domains. But the transformations from stripes to skyrmions are realized in large lamella and mostly not shown with reversed transformations.[35-42]

Here, we present a study on current-controlled skyrmion number in nanostructured $Fe_3Sn_2$ stripes. We use single pulsed currents of varying densities and durations within the range of 5 to 80 ns to achieve topological transformations between single dipolar-stabilized skyrmion chains and stripe domains. Our results demonstrate the reliable and reversible current-controlled skyrmion number in confined nanostripes, which could provide advancements in the development of topological spintronic devices.

## 2. Results and Discussions

### 2.1. Magnetic field-driven domain evolution of $Fe_3Sn_2$ nanostripe

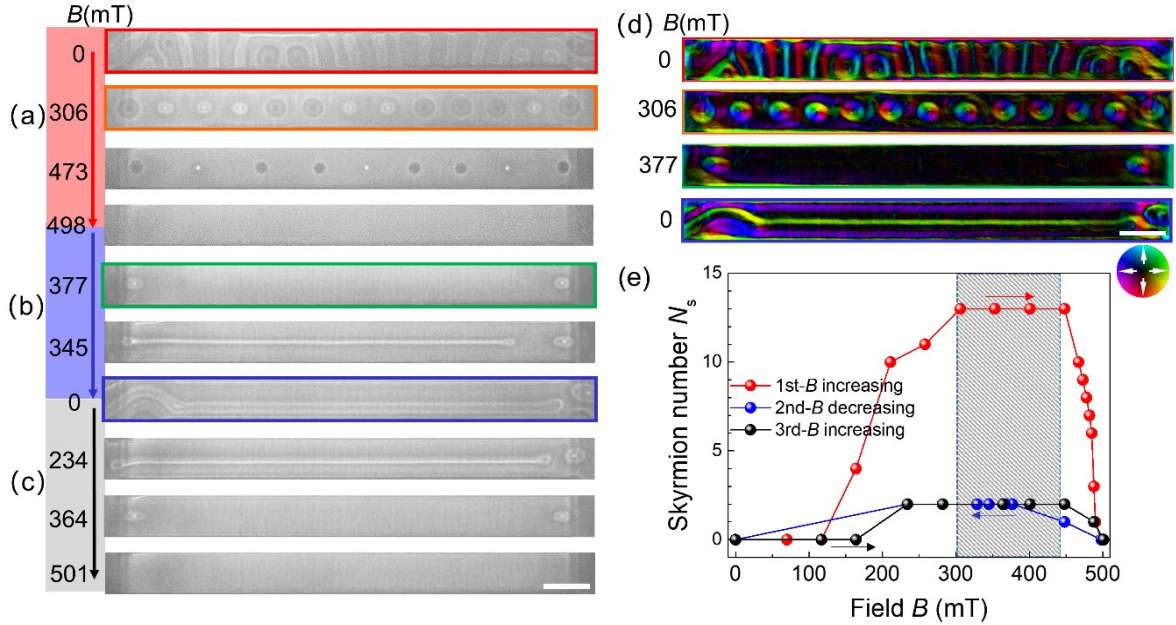

**Figure 1.** Static magnetic evolutions of the nanostructured $Fe_3Sn_2$ stripe. a) 1st, $B$-increasing process from the initial state of the as-fabricated device. b) 2nd, $B$-decreasing process from the saturated ferromagnetic (FM) state. c) 3rd, $B$-increasing process from zero fields. d) TIE analysis of the four Lorentz images marked by colorful rectangle frames in (a), (b), and (c). The color demonstrates the in-plane magnetic components according to the colorwheel. Dark contrast demonstrates zero in-plane magnetization. Scale bar, 500 nm. e) Skyrmion number $N_s$ as a function of the magnetic field $B$. The shadow region is the field range supporting two coexisting states, i.e. $N_s = 13$ and $N_s = 2$.

$Fe_3Sn_2$ is a centrosymmetric magnet that possesses a kagome lattice of Fe-Sn layers stacked along the $c$-axis.[43] Due to the absence of Dzyaloshinskii-Moriya interaction (DMI), dipolar-stabilized skyrmions (also called skyrmion bubbles or type-I bubbles) are present in this material instead of chiral skyrmions, as a result of the competition among exchange interaction, dipole-dipole interaction (DDI), uniaxial magnetic anisotropy, and Zeeman energy.[44, 45] We fabricated a $Fe_3Sn_2$ nanostripe (~ 5500 nm × 520 nm × 150 nm) using a focused ion beam (FIB) and placed it on an in-situ electrical chip. The resulting nanostructured sample was then observed in the Lorentz transmission electron microscopy (TEM) and in-situ current-induced dynamics was studied.

We first demonstrate the magnetic evolution of the nanostripe under an external magnetic field $B$ that is normal to the nanostripe. **Figure 1** shows the changes observed as the magnetic field increases and decreases. At $B = 0$ mT, multiple stripe domains with $q$-vector parallel to the length of the nanostripe are observed in the as-fabricated device. The stripe domains shrink and escape from the edge of the device to form skyrmions in the $B$-increasing process. At $B \sim 306$ mT, a single skyrmion chain with $N_s = 13$ is formed. Transport-of-intensity equation (TIE) analysis shows that the in-plane magnetization of dipolar skyrmions could rotate both clockwise and anti-clockwise (Figure 1d), which means that two different helicities exist simultaneously because of their degenerated energies.[46-52] With the further increase of $B$, the skyrmions gradually vanish and the FM states with no defocused Fresnel contrasts are observed for $B > 488$ mT. In the subsequent $B$-decreasing process from the saturation magnetic field, two skyrmions appear at the two ends of the nanostripe at $B \sim 377$ mT but no more new skyrmions are observed as the field continues to decrease. At $B \sim 345$ mT, the skyrmion at the left side expands and transforms into a long stripe domain, which is also called a stripy-skyrmion.[53] This stripy-skyrmion has the same skyrmion number as the round skyrmion observed at high fields.[53] As the magnetic field approaches zero, the stripe domain only becomes slightly curved and the skyrmion at the right side deforms into a stripe domain (Figure 1b). Finally, we explore the magnetic evolution in the subsequent $B$-increasing process and observe the emergence of at most two skyrmions at $B \sim 364$ mT (Figure 1c). Figure 1e shows the corresponding skyrmion number $N_s$ by cycling the magnetic field. Despite that $N_s = 13$ can be achieved in the 5500-nm long nanostripe, $N_s = 2$ at most can only be obtained in the subsequent cycles of the magnetic field. This is because the demagnetization field effect contributes to the stabilization of long stripy-skyrmions in the *B*-decreasing process from FM states. The experimental magnetic evolution is highly reproduced in our micromagnetic simulations (Supplemental Figure S1) based on measured material parameters. From the above analysis, we conclude that cycling the magnetic field alone cannot achieve a high skyrmion number. However, the possibility of tuning the skyrmion number at a fixed magnetic field using external stimuli such as currents is suggested since both the single skyrmion chain with $N_s = 13$ and the stripy-skyrmions with $N_s = 2$ are supported in the same field range of approximately 306 to 448 mT.

The skyrmion number $N_s$ is equal to the count of a skyrmion in the nanostripe, which has been verified through numerical simulations (Supplemental Figure S2). The simulated in-plane magnetization and Fresnel images for a skyrmion chain and stripe domain (Supplemental Figure S2) are all consistent with experimental observations well (Figure 1), indicating that simulations can be used to infer the skyrmion number $N_s$ in experiments. The skyrmion number $N_s$ is an integer for skyrmions with different helictites. Additionally, the $N_s$ of the long stripe domain is also equal to 1, which explains why it is referred to as stripy-skyrmion[53].

**2.2. Current-controlled the skyrmion number in the nanostructure.**

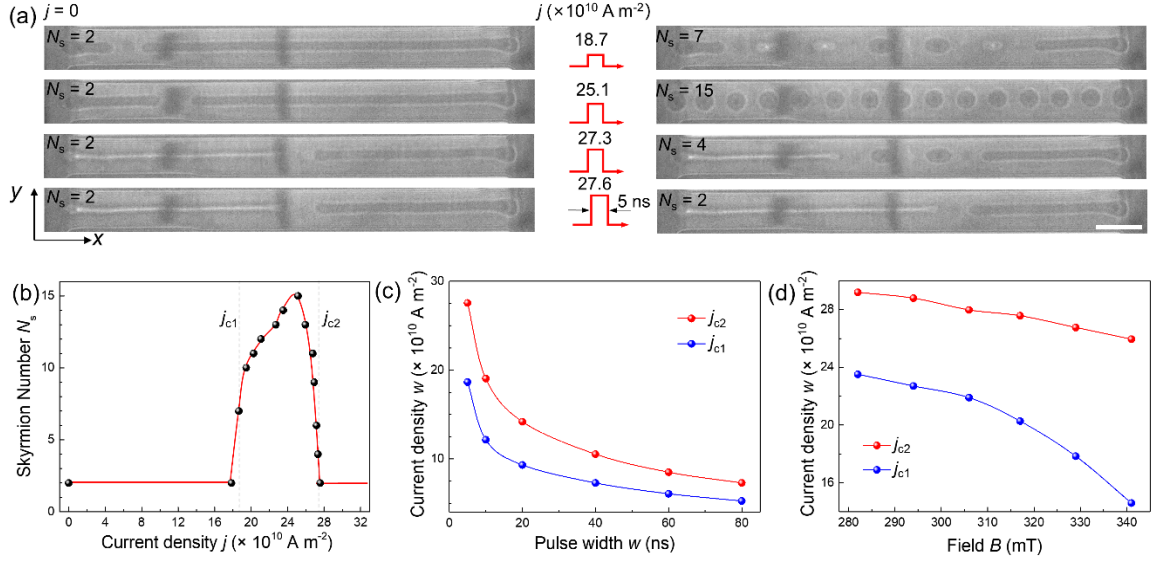

**Figure 2.** Current induced dynamics of stripy-skyrmions with $N_s = 2$. a) Typical Fresnel images of the magnetic textures before and after applying single 5-ns pulsed currents with different densities. The current is applied along the $+x$ axis. Defocus distance is -1 mm. The scale bar is 500 nm. b) Skyrmion number as a function of the current density $j$. The upper critical current density and lower critical current density are defined as $j_{c2}$ and $j_{c1}$, respectively. The current is applied along the $+x$ axis. $B \sim 317$ mT. c) Threshold current densities $j_{c1}$ and $j_{c2}$ as a function of pulse width $w$. d) Threshold current densities $j_{c1}$ and $j_{c2}$ as a function of the magnetic field $B$. $w = 5$ ns for (a), (b), and (d).

**Figure 2** illustrates the dynamic responses of stripy-skyrmions with $N_s = 2$ to single 5-ns pulsed currents along the $+x$ axis at a fixed magnetic field. For current density $j$ smaller than the threshold current density $j_{c1} \sim 18.7 \times 10^{10}$ A m$^{-2}$, there are no discernible dynamical changes. Once $j \geq j_{c1}$, the stripy-skyrmions break apart to form more skyrmions and $N_s$ is adjustable based on the current density $j$. $N_s$ first increases from 2 to 15 as $j$ increases from $17.8 \times 10^{10}$ A m$^{-2}$ to $25.1 \times 10^{10}$ A m$^{-2}$; subsequently, $N_s$ decreases from 15 to 2 as $j$ increases from $25.1 \times 10^{10}$ A m$^{-2}$ to $27.6 \times 10^{10}$ A m$^{-2}$ (Figures 2a, 2b, and Supplemental Figure S2). When $j$ exceeds a higher threshold current density $j_{c2}$, $N_s$ remains at 2. Current-induced effects include the Oersted field,[54, 55] where the Oersted field direction is dependent on the current orientation. However, reversing the current orientation cannot significantly impact the current-induced transition between the stripe and the skyrmion chain (Supplemental Figure S4). Hence, the Oersted field effect can be excluded as the dominant factor responsible for our observations.

The threshold current densities depend on both pulse duration $w$ and magnetic field $B$. By increasing the pulse duration from 5 to 80 ns, $j_{c2}$ decreases from 27.6 to 7.3 × $10^{10}$ A m$^{-2}$, and $j_{c1}$ decreases from 18.7 to 5.3 × $10^{10}$ A m$^{-2}$ (Figure 2c). By increasing the magnetic field $B$ from 282 to 341 mT, $j_{c2}$ decreases from 29.2 to 26.0 × $10^{10}$ A m$^{-2}$ and $j_{c1}$ decreases from 23.5 to 14.6 × $10^{10}$ A m$^{-2}$ (Figure 2d).

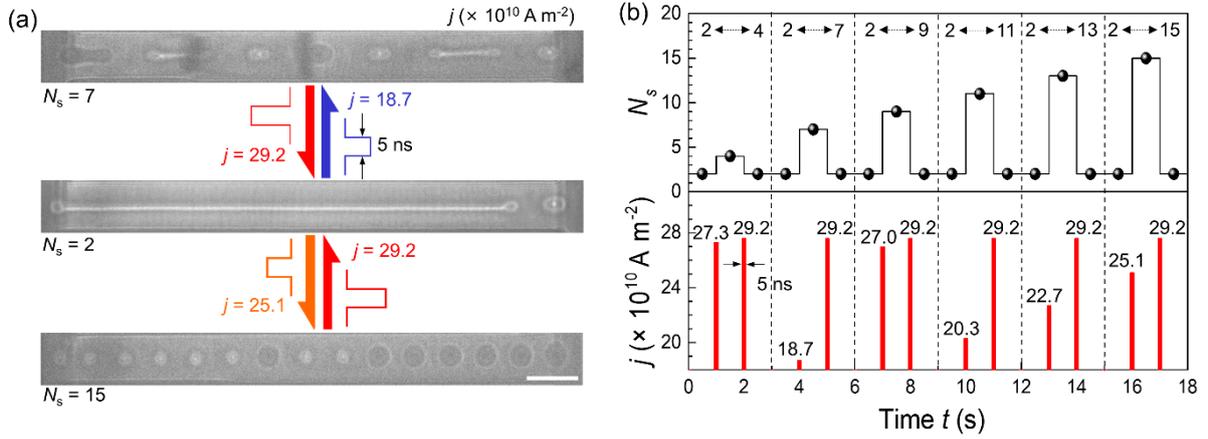

**Figure 3.** Reliable manipulation of the skyrmion number in the nanostripe by the current. a) Representative reversal topological transformation between a single skyrmion chain and stripe-skyrmions at $B \sim 317$ mT. Transformations between $N_s = 2$ and $N_s = 7$ states are achieved by cycling two discrete pulsed current densities with $j = 29.2 \times 10^{10}$ and $18.7 \times 10^{10}$ A m$^{-2}$, respectively. Transformations between $N_s = 2$ and $N_s = 15$ states are achieved by cycling two discrete pulsed current densities with $j = 29.2 \times 10^{10}$ and $25.1 \times 10^{10}$ A m$^{-2}$, respectively. Scale bar, 500 nm. Defocus distance, -1 mm. b) Skyrmion numbers $N_s$ induced by different current densities of pulse cycles. Each cycle includes a lower current density and a higher current density. Pulse duration $w = 5$ ns.

The results discussed in Figure 2 suggest that the skyrmion number $N_s$ can be tuned through transitions between a single skyrmion chain and stripy-skyrmions by applying two discrete current densities. Pulsed currents of 5-ns width with varying densities were injected alternately while keeping the magnetic field constant (**Figure 3**a). When a low-density pulsed current ($j = 18.7 \times 10^{10}$ A m$^{-2}$) is applied to the initial stripy-skyrmions with $N_s = 2$, a single skyrmion chain with $N_s = 7$ generates. When a high-density pulsed current ($j = 29.2 \times 10^{10}$ A m$^{-2}$) is injected into the single skyrmion chain, the single skyrmion chain transforms to two stripy-skyrmions with $N_s = 2$ (Figure 3a). Similarly, by switching the current density between 25.1 and $29.2 \times 10^{10}$ A m$^{-2}$, reversible switching between $N_s = 15$ and $N_s = 2$ states is demonstrated (Figure 3a). Figure 3b illustrates representative examples of reversible control of skyrmion numbers by injecting different pairs of low and high currents. Convenient acquisition of $N_s = 4, 7, 9, 11, 13$, and 15 from an initial state of $N_s = 2$ can be achieved through this method (Supplementary Videos S1-S6).

Since the initial state of $N_s = 2$ can always be obtained by a single pulsed current with a density $j > j_{c2}$, we propose that topological transformations between any two skyrmion numbers can be achieved through a double-pulsed method (Supplemental Figure S5 and Videos S7-S9). The double-pulsed method involves initially applying a high current with $j > j_{c2}$ to initialize the state with $N_s = 2$. Then, a second pulsed current is applied to achieve the desired skyrmion number state according to the results in Figure 3. This approach could enable the realization of a wide range of skyrmion numbers by using different current densities, portending promising prospects for multi-bit memory and computing devices in the future.

**2.3. Physical origins of current-controlled skyrmion number.**

The physical mechanisms behind current-controlled skyrmion number are investigated via micromagnetic simulations. The threshold current densities are strongly dependent on the pulse duration $w$, with an increase in $w$ leading to a decrease in threshold current densities
9

(Figure 2c), suggesting that the Joule thermal heating effects could be responsible for our results. Previous research has established that a rising temperature can be induced by currents, for example, a 100-ns pulsed current with $j \sim 3.4 \times 10^{10}$ A m$^{-2}$ results in a rising temperature of ~ 180 K.[56] The coexisted magnetic states with different skyrmion numbers at a fixed field (Figure 1e), such as the stripy-skyrmion state with $N_s = 2$ and the skyrmion chain with $N_s = 13$, suggest an energy barrier between them. While both $N_s = 2$ and $N_s = 13$ can be supported over a wide field range from ~ 306 to ~ 448 mT, current-controlled skyrmion number between $N_s = 2$ and 13 can be realized only in the low field range of ~ 306 to ~ 341 mT. In the low-field range, the stripe state appears to be more stable than the skyrmion chain, as shown in **Figure 4**a. However, during the temperature-rising process, the stripe state becomes more unstable than the skyrmion chain because of the reduction of the saturation field at high temperatures. Once the thermal fluctuation energy $E_{therm}$ is comparable to the energy barrier between the two states, the stripe state prefers to break to form various skyrmions (Supplemental Video S10). Figure 4b shows the magnetic evolution of a stripe considering thermal fluctuation energy. In the confined nanostripe at a fixed field of $B = 300$ mT, $N_s$ increases from 1 to 8 in 50 ns once a thermal fluctuation field is applied and agrees with the schematic energy profile in Figure 4a. Figure 4c shows that skyrmion number varies with different thermal fluctuation fields. As the temperature increases, the stripe state with skyrmion number $N_s = 1$ remains unchanged at first. When the temperature reaches a certain threshold value, the skyrmion number $N_s$ begins to increase from 1 to 13. Then the $N_s$ decreases to 0 as the temperature keeps going up and $N_s = 0$ is maintained for higher temperature. The relationship between the skyrmion number and the temperature shows a very similar tendency as Figure 2b and Supplemental Figure S3, indicating that the skyrmion number controlled by current densities results from the Joule thermal effect.

Despite the fact that the saturation magnetization $M_s$ and anisotropy constant $K_u$ are not changed when considering thermal fluctuation fields in simulations, the effective saturation magnetization $M_{\text{s-eff}}$ and anisotropy constant $K_{\text{u-eff}}$, which reflect the realistic magnetic parameters in experiments, decrease with increasing temperature (Supplemental Figure S6).

We have demonstrated the stripe-to-skyrmion transformation through current-induced thermal heating effects. However, it is important to note that thermal effects from high-density currents may differ significantly. One prominent thermal effect is the reduction of saturated magnetization, resulting in a decrease in the saturated magnetization field. Despite the field is fixed during the injection of current, the saturated magnetization field can be smaller than the fixed field due to a high-temperature rise associated with high-density currents, resulting in only a stable FM state (Figure 4c), which is indexed as stage-I of thermal heating as illustrated in **Figure 5**a. Once the pulsed current is turned off, the temperature of the device decreases back to room temperature, leading to stage II of the magnetization recovery process. Our experiments have shown that from the initial FM state, high $N_s$ states are not readily accessible (Figure 1). This is due to the demagnetization field effect that promotes the stabilization of long stripe domains, as demonstrated in Figures 5b and 5c (Supplemental Video S11). Consequently, in stage II of the magnetization recovery process, few skyrmions with low $N_s$ states are more easily achieved from the initial FM state that matches the energy profiles illustrated in Figure 5a.

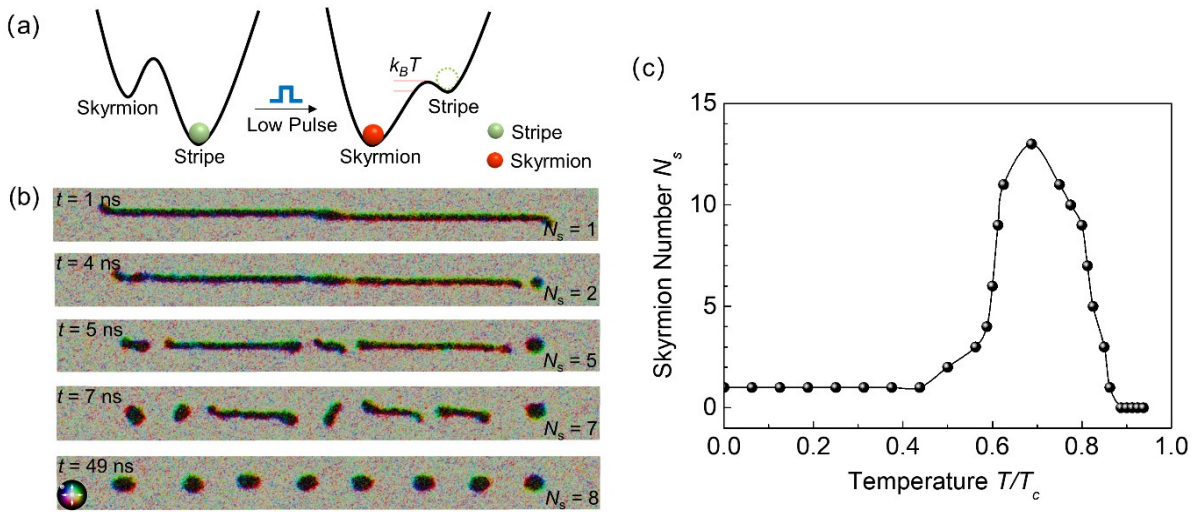

**Figure 4.** Simulated transformation from the stripe domain to the single skyrmion chain through thermal effects of a low current. a) Schematic diagram of energy density profiles for skyrmions transforming to stripe domains. The green sphere represents the stripy-skyrmion state and the red one represents the skyrmion state. b) Transition from the stripe to the skyrmion chain at $T/T_c$ = 0.61. $T_c$ is the Curie temperature. The five images are recorded at $t$ =1, 4, 5, 7, and 49 ns, with the skyrmion number $N_s$ = 1, 2, 5, 7, and 8. The color wheel indicates the orientation of magnetization. Dark and white contrasts represent out-of-plane down and up magnetizations, respectively. The length, width, and thickness of the nanostripe in simulations are 5000, 500, and 150 nm, respectively. c) Skyrmion number $N_s$ as a function of normalized temperature $T/T_c$.

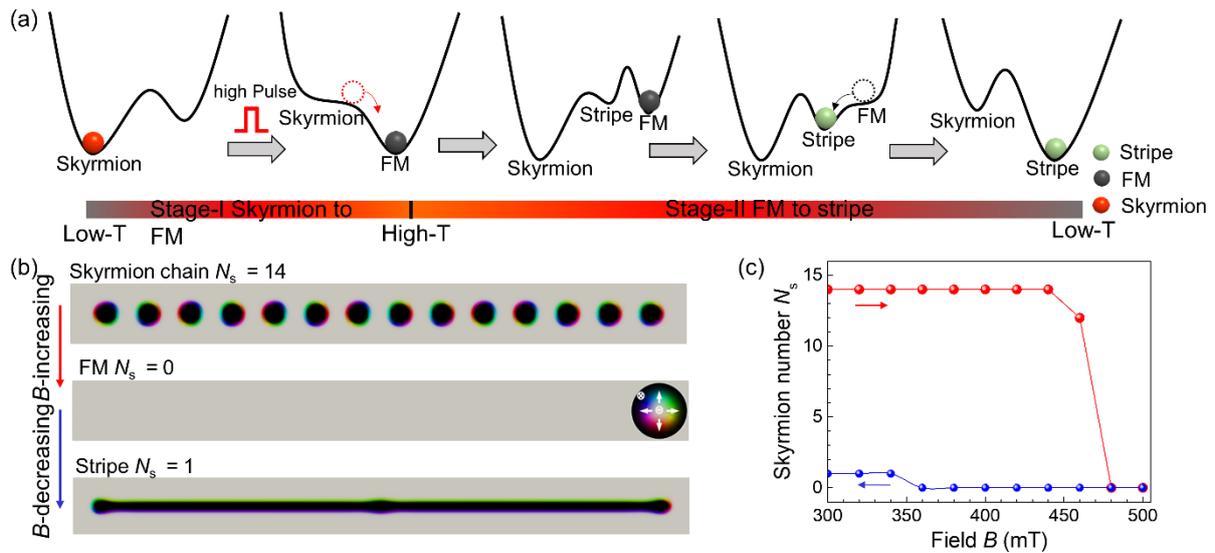

**Figure 5.** Transformation from the single skyrmion chain to the stripe through the thermal effects of a high current. a) Schematic energy profiles for the stripe-to-skyrmion transformation. The green, black, and red sphere represents the stripe state, FM state, and skyrmion state, respectively. b) Representative magnetic states by varying magnetic fields in zero-temperature simulations. The color wheel indicates the orientation of magnetization. Dark and white contrasts represent out-of-plane down and up magnetizations, respectively. c) Skyrmion number $N_s$ as a function of magnetic field $B$.

It should be noted that the skyrmion state in $Fe_3Sn_2$ at room temperature is quite thermally stable, due to the remarkably high Curie temperature (~680 K), as well as the excellent chemical and physical stability of the $Fe_3Sn_2$ crystal.[43, 57-59] We test that the skyrmion state remains unchanged for over 7 days (Supplemental Figure S7) under the same physical conditions.

## 3. Conclusions

In summary, we have demonstrated that the manipulation of the skyrmion number can be achieved flexibly in confined nanostripes using pulsed currents. In combination with micromagnetic simulations, the current-induced dynamics are explained by thermal effects and magnetic hysteresis, which have entirely different phenomena for low and high currents. The stripe-to-skyrmion transformation can occur through low-density currents due to thermal fluctuation energy. Conversely, the skyrmion-to-stripe transformation through high-density currents is due to the hysteresis process from FM states. Our findings highlight the potential of using nanosecond-pulsed currents to reversibly and reliably manipulate the skyrmion number in nanostripes. Furthermore, the voltage-controlled skyrmion number is achieved through the strain change transferred by a specially-designed ferroelectric PMN-PT substrate[34]. Here, the current-controlled skyrmion number can be realized by simply applying electrodes on the two ends of nanostripes without any additional geometries. This straightforward design holds the potential for promoting the development of fast and reliable topological spintronic devices.

## 4. Methods

*Sample Preparation*: $Fe_3Sn_2$ single crystal was grown by chemical vapor transport method from stoichiometric Fe (>99.9%) and Sn (>99.9%) mixture. The 150-nm-thick nanostripe of $Fe_3Sn_2$, fixed on an in-situ electrical chip, was fabricated via a standard lift-out method from the bulk by using a focused ion beam instrument (Helios Nanolab 600i, FEI).[44]

*TEM Measurements*: Fresnel magnetic images were recorded by a TEM instrument (Talos F200X, FEI), operated at 200 kV and Lorentz mode.[46] The perpendicular magnetic field varies by adjusting the object current. The pulsed currents were provided by using a voltage source (AVR-E3-B-PN-AC22, Avtech Electrosystems). The experiments were all performed at room temperature.

*Micromagnetic Simulations*: Micromagnetic Simulations were performed by using a GPU-accelerated micromagnetic simulation program, Mumax3.[60] The exchange interactions, uniaxial magnetic anisotropy, Zeeman energy, and dipole-dipole interactions were considered in the simulations. Magnetic parameters were set based on the $Fe_3Sn_2$ material with A = 8.25 pJ/m, $K_u$ = 54.5 kJ/m$^3$, and saturation magnetization $M_s$ = 622.7 kA/m[44, 45]. The cell size was set to 5 × 5 × 5 nm$^3$. The equilibrium spin configurations were obtained by using the conjugate-gradient method. The finite temperature was provided by the thermal fluctuation filed $\vec{B}_{\text{therm}} = \vec{\eta}(\text{step})\sqrt{\frac{2\mu_0 \alpha k_B T}{B_{\text{sat}} \gamma_{\text{LL}} \Delta V \Delta t}}$, where $\alpha$, $k_B$, $T$, $B_{\text{sat}}$, $\gamma_{\text{LL}}$, $\Delta V$, and $\Delta t$ are the damping factor, Boltzmann constant, temperature, saturation magnetization, gyromagnetic ratio, cell volume, and time step, respectively. $\vec{\eta}(\text{step})$ is a random vector changed after every time step.[61]

**Supporting Information**

Supporting Information is available from the Wiley Online Library or from the author.

**Acknowledgments**

This work is supported by the National Key R&D Program of China, Grant No. 2022YFA1403603 and 2021YFA1600200; the Natural Science Foundation of China, Grants No. 12174396, 12104123, and 11974021; Natural Science Project of Colleges and

Universities in Anhui Province, Grant No. 2022AH030011; the Strategic Priority Research Program of Chinese Academy of Sciences, Grant No. XDB33030100; and Innovation Program for Quantum Science and Technology, Grant No. 2021ZD0302802.**Author Contributions**

J. J. and J. T. contributed equally to this work. H. D. and J. T. supervised the project. J. T. conceived the idea and designed the experiments. Y. Wang, J. L., and Y. X. synthesized the $Fe_3Sn_2$ single crystals. J. J. fabricated the $Fe_3Sn_2$ nanostripes. J. J., J. T., and Y. Wu performed the TEM measurements. J. T. performed the simulations with the help of Q. Z.. J. T., J. J., and H. D. wrote the manuscript with input from all authors. All authors discussed the results and contributed to the manuscript.

**Conflict of Interest:**

The authors declare no competing financial interest.

# Supporting Information
# Current-Controlled Skyrmion Number in Confined Ferromagnetic Nanostripes

*Jialiang Jiang[#], Jin Tang[#]\*, Yaodong Wu, Qi Zhang, Yihao Wang, Junbo Li, Yimin Xiong, Lingyao Kong, Shouguo Wang, Mingliang Tian, and Haifeng Du\**

Jialiang Jiang, Yaodong Wu, Yihao Wang, Junbo Li, Mingliang Tian, Haifeng Du
Anhui Province Key Laboratory of Condensed Matter Physics at Extreme Conditions, High Magnetic Field Laboratory, Hefei Institutes of Physical Science, Anhui, Chinese Academy of Sciences, Hefei 230031, China
Email: duhf@hmfl.ac.cn

Jialiang Jiang, Yihao Wang, Junbo Li
University of Science and Technology of China, Hefei 230026, China

Jin Tang, Qi Zhang, Yimin Xiong, Lingyao Kong, Mingliang Tian
School of Physics and Optoelectronics Engineering Science, Anhui University, Hefei, 230601, China
Email: jintang@ahu.edu.cn

Shouguo Wang
School of Materials Science and Engineering, Anhui University, Hefei 230601, China

[#]Jin Tang and Jialiang Jiang contributed equally to this work.

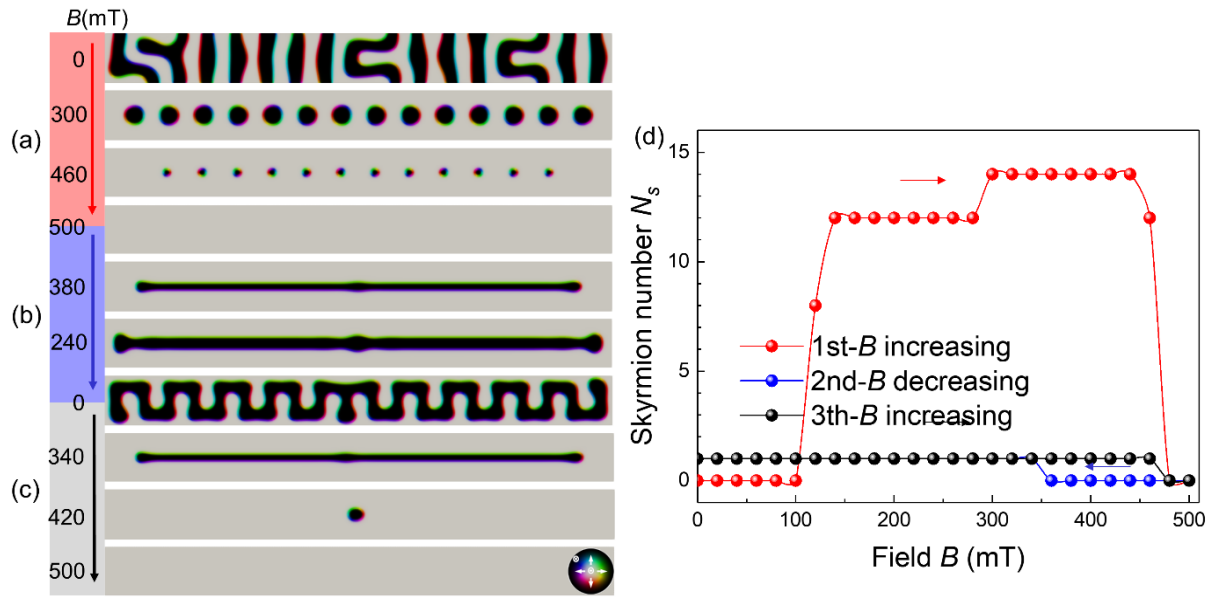

**Figure S1.** Zero-temperature simulation for magnetic evolutions of the nanostructured $Fe_3Sn_2$ sample. a) 1$^{st}$, $B$-increasing process from the multiple stripes state. b) 2$^{nd}$, $B$-decreasing process from the saturated FM state. c) 3$^{rd}$, $B$-increasing process from zero fields. The color wheel indicates the orientation of magnetization. Dark and white contrasts represent out-of-plane down and up magnetizations, respectively. Scale bar, 500 nm. d) Skyrmion number $N_s$ as a function of the magnetic field $B$.

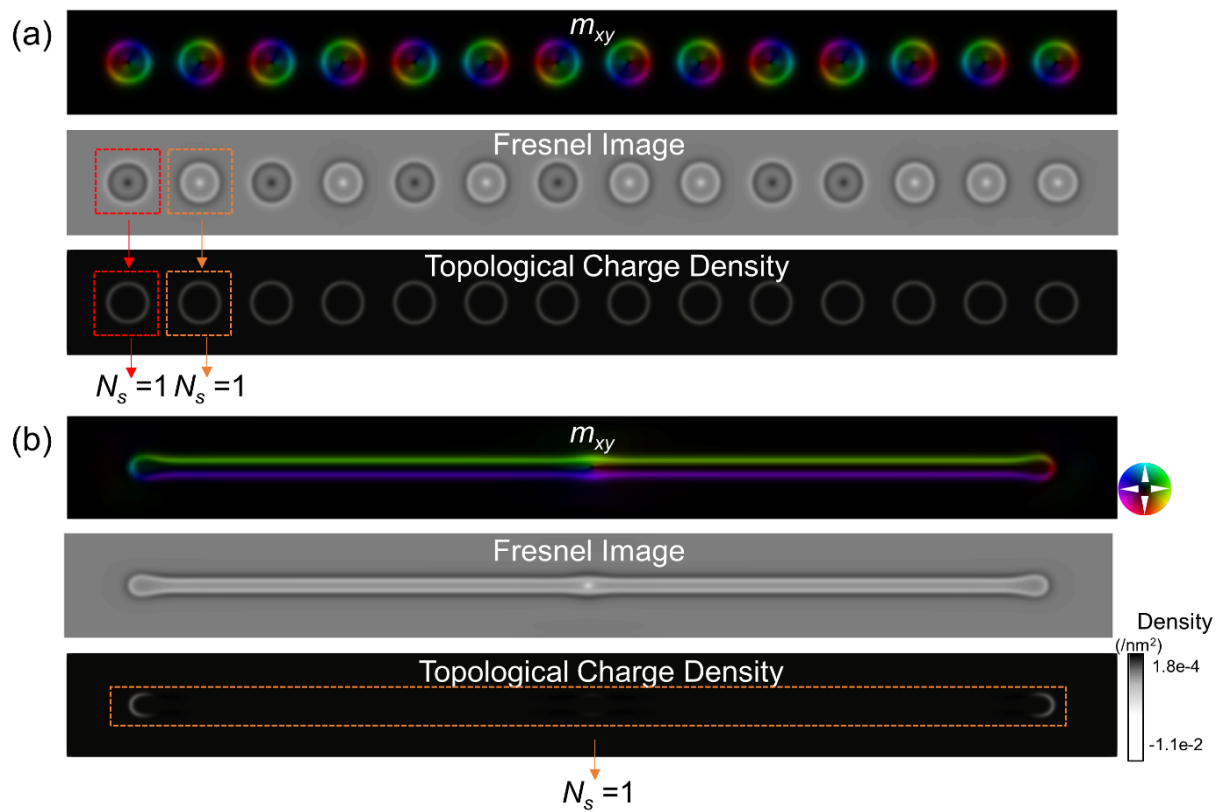

**Figure S2.** In-plane magnetization distribution, corresponding fresnel image, and spatial distribution of topological charge density of a skyrmion chain with $N_s = 14$ (a) and a stripy-skyrmion (b) with $N_s = 1$. The integrated topological charges of skyrmions integrated over three marked regions are all equal to 1. For $m_{xy}$ images, the color demonstrates the in-plane magnetic components according to the colorwheel and dark represents zero in-plane magnetization. For topological charge density images, the grayscale bar indicates topological charge density per square nanometer.

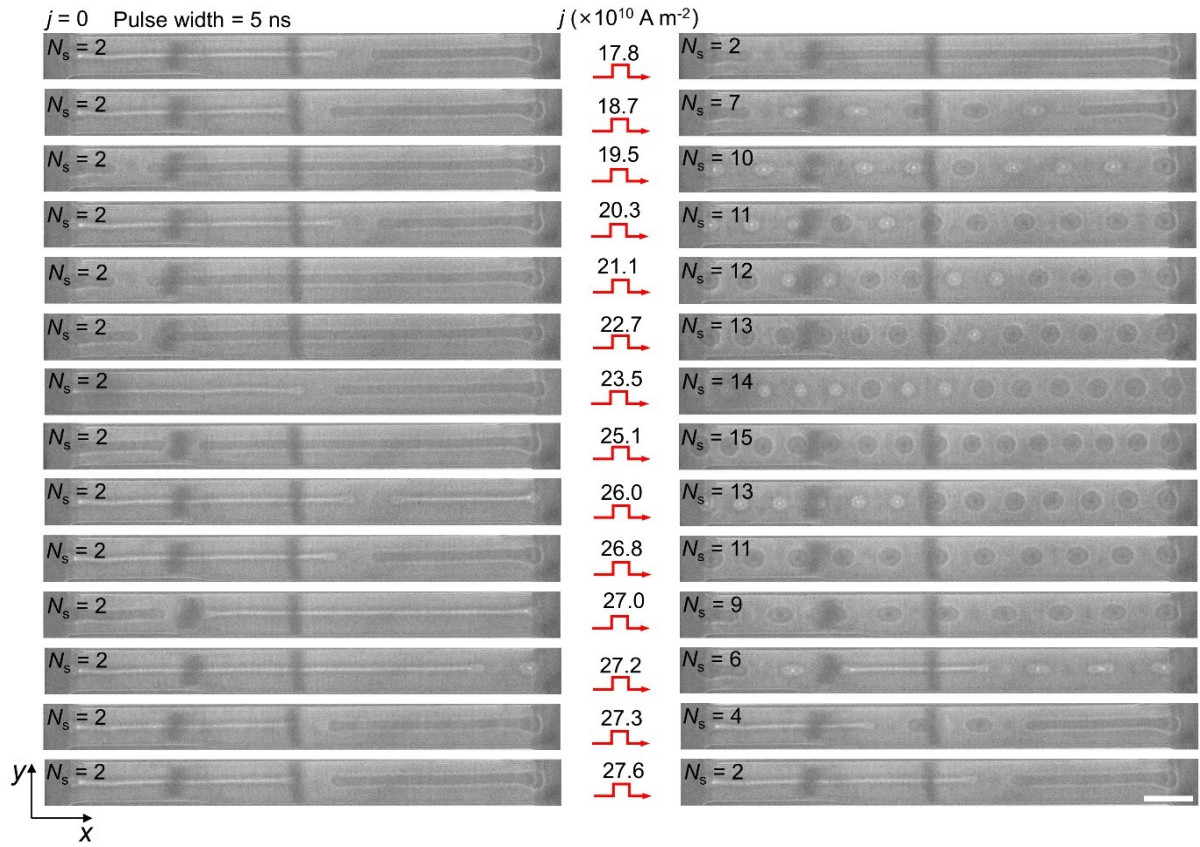

**Figure S3.** Current induced dynamics of stripy-skyrmions with $N_s = 2$. All the Fresnel images of the magnetic textures before and after applying single 5-ns pulsed currents with different densities are given. The current is applied along the $+x$ axis. Defocus distance is -1 mm. The scale bar is 500 nm. The current is applied along the $+x$ axis. $B \sim 317$ mT.

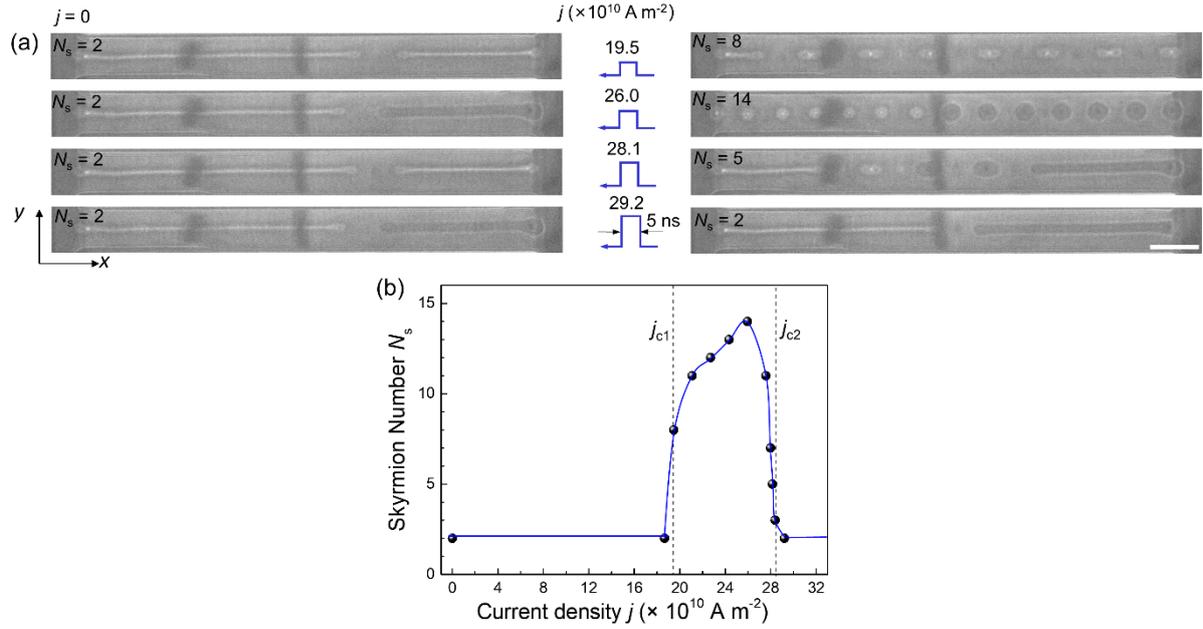

**Figure S4.** Current with opposite direction induced dynamics of stripy-skyrmions with $N_s = 2$. a) Typical Fresnel images of the magnetic textures before and after applying single 5-ns pulsed currents with different densities. The current is applied along the -x axis. Defocus distance is -1 mm. The scale bar is 500 nm. b) Skyrmion number as a function of the current density $j$. The upper critical current density and lower critical current density are defined as $j_{c2}$ and $j_{c1}$. The current is applied along the -x axis. $B \sim 317$ mT.

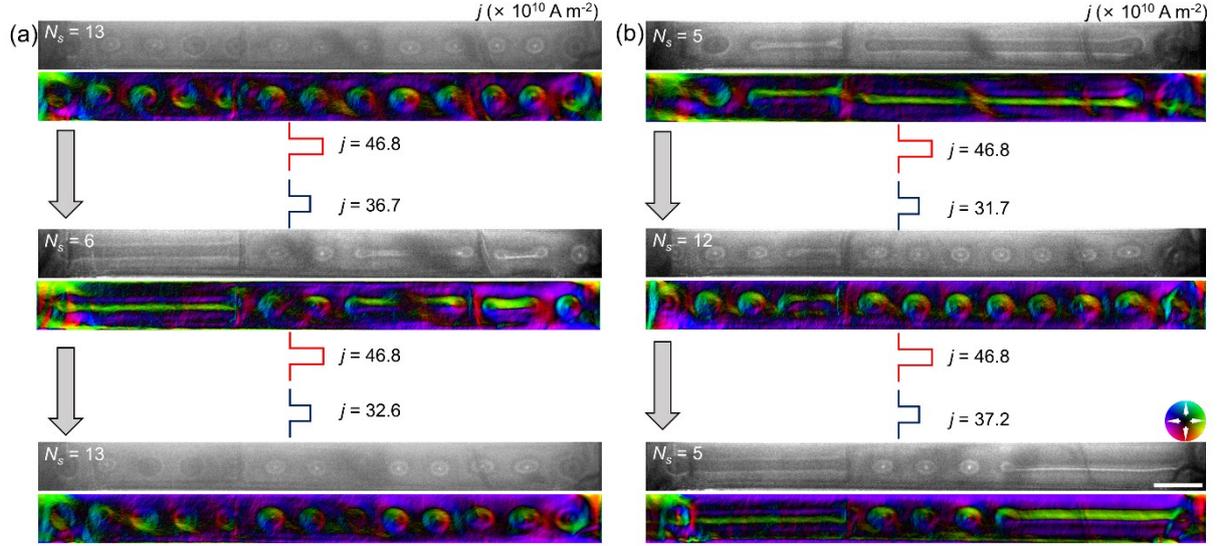

**Figure S5.** Representative reversed topological transformation between any two $N_s$ states by using a double-pulsed method. a) Transformations between $N_s = 13$ and $N_s = 6$ states are achieved by applying a double-pulsed current density with $j = 46.8 \times 10^{10}$ A m$^{-2}$ and $36.7 \times 10^{10}$ A m$^{-2}$ in turn. By using double-pulsed current densities with $j = 46.8 \times 10^{10}$ A m$^{-2}$ and $32.6 \times 10^{10}$ A m$^{-2}$, the $N_s = 6$ state can be transformed back to the $N_s = 13$ state. B) Transformations between $N_s = 5$ and $N_s = 12$ states are achieved by applying a double-pulsed current density with $j = 46.8 \times 10^{10}$ A m$^{-2}$ and $31.7 \times 10^{10}$ A m$^{-2}$ in turn. By using a double-pulsed current density with $j = 46.8 \times 10^{10}$ A m$^{-2}$ and $37.2 \times 10^{10}$ A m$^{-2}$, the $N_s = 12$ state can be transformed back to the $N_s = 5$ state. The color demonstrates the in-plane magnetic components according to the colorwheel. Dark contrast demonstrates zero in-plane magnetization. Scale bar, 500 nm. Defocus distance, -1 mm. $B \sim 294$ mT.

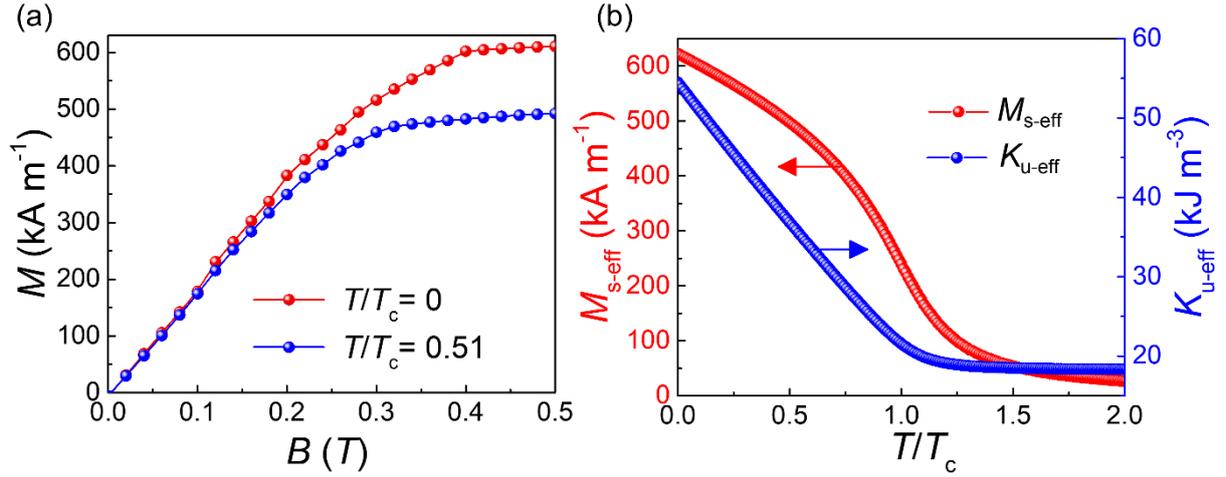

**Figure S6.** a) Magnetization $M$ as a function of magnetic field $B$ for $T/T_c = 0.51$ and 0. The lower $M_{\text{s-eff}}$ for $T/T_c = 0.51$ is attributed to the spin fluctuations around the field direction induced by the thermal effects. The spin fluctuations result in a decrease in the average magnetization over time, which is consistent with experimental observations. b) Effective saturation magnetization $M_{\text{s-eff}}$ and anisotropy constant $K_{\text{u-eff}}$ as a function of thermal fluctuation temperature $T/T_c$. The anisotropic energy is defined as $\varepsilon_{\text{ani}} = -K_u(\mathbf{m} \cdot \mathbf{u})^2$, with $\mathbf{u} = (0,0,1)$. At zero temperature, the ferromagnetic (FM) state aligns with the external field $\mathbf{B} = (0,0,B_z)$, resulting in a net magnetization $\bar{\mathbf{M}} = M_s|\mathbf{m}| = M_s(0,0,1)$ associated with the anisotropic energy $\varepsilon_{\text{ani}} = -K_u$. Under a finite temperature $T$, accounting for fluctuations, the net magnetization should be considered as $\bar{\mathbf{M}}(T) = M_s < \mathbf{m}(T) >$. The average magnetization $< \mathbf{m}(T) >$ is always smaller than 1 due to thermal fluctuation. Thus, the effective anisotropic energy, which is estimated as $\varepsilon_{\text{ani-eff}}(T_1) = -K_u < \mathbf{m}_z(T)^2 >$, is also lower than that at zero temperature. If we introduce the effective magnetization $M_{s-\text{eff}}$ and effective anisotropy constant $K_{u-\text{eff}}$ to describe this temperature-dependent behavior, then:

$$M_{s-\text{eff}} = M_s < \mathbf{m}_z(T) >$$
$$K_{u-\text{eff}} = K_u < \mathbf{m}_z(T)^2 >$$

The average magnetization $< \mathbf{m}_z(T) >$ is obtained by applying a saturation magnetic field along the [0, 0, 1] axis. $M_s = 622.7$ kA/m and $K_u = 54.5$ kJ/m³ for all temperatures.

Initial state
$N_s = 5$

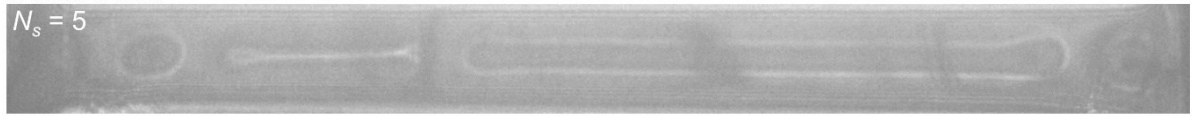

After 7 days
$N_s = 5$

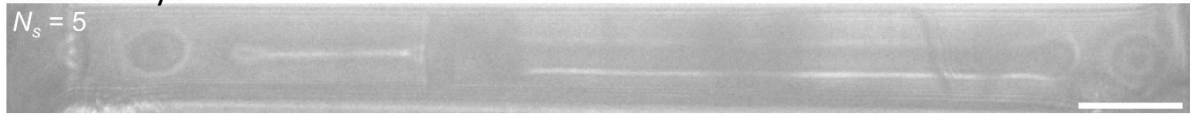

**Figure S7.** Thermal stability of the skyrmions state in Fe$_3$Sn$_2$ nanostructured sample. $B = \sim 294$ mT at room temperature. Scale bar, 500 nm. Defocus distance, -1 mm.

**Supplemental Movie captions:**

**Supplemental Video S1.** Transformations between $N_s = 2$ and $N_s = 4$ states are achieved by cycling two discrete pulsed current densities with $j = 29.2 \times 10^{10}$ A m$^{-2}$ and $27.3 \times 10^{10}$ A m$^{-2}$. Pulse duration $w = 5$ ns. $B = \sim 317$ mT.

**Supplemental Video S2.** Transformations between $N_s = 2$ and $N_s = 7$ states are achieved by cycling two discrete pulsed current densities with $j = 29.2 \times 10^{10}$ A m$^{-2}$ and $18.7 \times 10^{10}$ A m$^{-2}$. Pulse duration $w = 5$ ns. $B = \sim 317$ mT.

**Supplemental Video S3.** Transformations between $N_s = 2$ and $N_s = 9$ states are achieved by cycling two discrete pulsed current densities with $j = 29.2 \times 10^{10}$ A m$^{-2}$ and $27.0 \times 10^{10}$ A m$^{-2}$. Pulse duration $w = 5$ ns. $B = \sim 317$ mT.

**Supplemental Video S4.** Transformations between $N_s = 2$ and $N_s = 11$ states are achieved by cycling two discrete pulsed current densities with $j = 29.2 \times 10^{10}$ A m$^{-2}$ and $20.3 \times 10^{10}$ A m$^{-2}$. Pulse duration $w = 5$ ns. $B = \sim 317$ mT.

**Supplemental Video S5.** Transformations between $N_s = 2$ and $N_s = 13$ states are achieved by cycling two discrete pulsed current densities with $j = 29.2 \times 10^{10}$ A m$^{-2}$ and $22.7 \times 10^{10}$ A m$^{-2}$. Pulse duration $w = 5$ ns. $B = \sim 317$ mT.

**Supplemental Video S6.** Transformations between $N_s = 2$ and $N_s = 15$ states are achieved by cycling two discrete pulsed current densities with $j = 29.2 \times 10^{10}$ A m$^{-2}$ and $25.1 \times 10^{10}$ A m$^{-2}$. Pulse duration $w = 5$ ns. $B = \sim 317$ mT.

**Supplemental Video S7.** Transformations between $N_s = 13$ and $N_s = 6$ states are achieved reversely by using a double-pulsed method. Pulse duration $w = 5$ ns. $B = \sim 294$ mT.

**Supplemental Video S8.** Transformations between $N_s = 12$ and $N_s = 5$ states are achieved reversely by using a double-pulsed method. Pulse duration $w = 5$ ns. $B = \sim 294$ mT.

**Supplemental Video S9.** Transformations between $N_s = 12$ and $N_s = 4$ states are achieved reversely by using a double-pulsed method. Pulse duration $w = 5$ ns. $B = \sim 294$ mT.

**Supplemental Video S10.** Simulated transformation process from the stripe domain to the single skyrmion chain under the activation of a thermal fluctuation field $T/T_c = 0.61$. $B = 300$ mT. The size of the simulated sample is $5000 \times 500 \times 150$ nm$^3$.

**Supplemental Video S11.** Simulated magnetic evolutions of the nanostructured Fe$_3$Sn$_2$ sample. The size of the simulated sample is $5000 \times 500 \times 150$ nm$^3$.